\documentclass[10pt,twocolumn,letterpaper]{article}

\usepackage[margin=0.75in]{geometry}
\usepackage{titlesec}
\titlespacing*{\section}{0pt}{5pt}{0pt}
\titlespacing*{\subsection}{0pt}{2pt}{0pt}

\usepackage{amsmath}
\usepackage{amssymb}

\usepackage[table]{xcolor} 
\usepackage{booktabs}      

\usepackage{algorithm}
\usepackage{algpseudocode}
\usepackage{enumitem}
\usepackage{graphicx}
\usepackage{siunitx}
\usepackage{subcaption} 
\usepackage{hyperref}
\usepackage{authblk}
\usepackage{balance}

\newcommand{\ie}{\emph{i.e.}, }
\newcommand{\eg}{\emph{e.g.}, }

\providecommand{\keywords}[1]{\textbf{\textit{Keywords---}} #1}

\title{Thin-Client Interactive Gaussian Adaptive Streaming over HTTP/3}

\author[1]{Emanuele Artioli}
\author[1]{Philipp Fößl}
\author[2]{Daniele Lorenzi}
\author[1]{Farzad Tashtarian}
\author[3]{Mahdi Dolati}
\author[4]{Cheng-Hsin Hsu}
\author[1]{Christian Timmerer}

\affil[1]{Alpen-Adria-Universität, Klagenfurt, Austria}
\affil[2]{Bitmovin Inc., Klagenfurt, Austria}
\affil[3]{Sharif University of Technology, Tehran, Iran}
\affil[4]{National Tsing Hua University, Taiwan}

\date{}

\begin{document}
\sloppy
\maketitle

\begin{abstract}
Recent advancements in 3D Gaussian Splatting (3DGS) have enabled photorealistic rendering of complex scenes, yet widespread adoption on mobile and Extended Reality (XR) devices is hindered by substantial computational and bandwidth requirements. While existing solutions often focus on model compression for client-side rendering, they still demand significant GPU power, limiting applicability on resource-constrained hardware. We propose TIGAS (Thin-client Interactive Gaussian Adaptive Streaming), a remote rendering framework offloading rasterization to a backend. To bypass the prohibitive latencies connected to fluctuating network conditions, TIGAS streams view-dependent 2D projections to a lightweight web client over QUIC, minimizing head-of-line (HoL) blocking. A dedicated ABR algorithm adapts rendering quality to fluctuating network conditions, maintaining motion-to-photon latency within strict 6DoF interactive constraints. Furthermore, we discuss the integration of an experimental WebGPU super-resolution pipeline to analyze the trade-offs between perceptual quality enhancements and thin-client processing bottlenecks. We extensively evaluate TIGAS across multi-continental environments using 14 3DGS models and real 6DoF EyeNavGS movement traces. Powered by a backend rendering frames in under 10 milliseconds, TIGAS maintains latency within interactive thresholds while achieving an average SSIM of 0.88, serving both as a robust testbed for 3DGS streaming research and a capable delivery system. The source code is available at: \hyperlink{https://github.com/Rekenar/GaussianAdaptiveStreamer}{https://github.com/Rekenar/GaussianAdaptiveStreamer}.
\end{abstract}


\section{Introduction}
The proliferation of Extended Reality (XR) devices has accelerated the demand for immersive, photorealistic 3D content on commodity hardware. Unlike 2D video, these applications require six Degrees of Freedom (6DoF) navigation, allowing users to explore virtual environments from arbitrary viewpoints. To meet visual fidelity expectations, 3D Gaussian Splatting (3DGS)~\cite{Kerbl2023} has emerged as a state-of-the-art rendering technique. By representing scenes as millions of anisotropic 3D Gaussians, 3DGS achieves photorealism comparable to Neural Radiance Fields (NeRFs)~\cite{mildenhall2020nerf} but with faster real-time rendering speeds~\cite{TVCG_3DGS_Survey}. 

However, widespread adoption of 3DGS on mobile and commodity hardware faces a ``delivery gap.'' High-fidelity 3DGS models are data-intensive, often requiring gigabytes of storage for complex scenes, which translates to prohibitive bandwidth requirements for real-time streaming~\cite{SGSS}. Furthermore, while 3DGS rendering is faster than NeRFs, it remains computationally expensive. Rendering millions of splats imposes a heavy load on the GPU, often exceeding the thermal and power budgets of mobile devices. Recent benchmarks indicate that even capable edge platforms, such as the NVIDIA Jetson AGX Orin, struggle to maintain consistent \qty{90}{fps} seamless motion-sickness-free immersion in complex interactive scenes~\cite{Wei2025}. 

Existing 3DGS streaming solutions largely focus on \emph{native streaming}, where raw Gaussian data (or a compressed version thereof) is transmitted to the client for local rendering~\cite{sun20253dgsabr, shi2025lapisgs}. While effective on powerful desktops, this approach excludes resource-constrained ``thin'' clients (\eg smartphones, smart TVs, basic laptops) that lack the dedicated GPU hardware required for high-frequency rasterization. Furthermore, besides these computational constraints, native streaming requires either downloading massive amounts of data before the interactive session begins, causing prohibitive startup delays, or receiving only a partial subset of the model, resulting in consequent quality degradation. 

On the other hand, attempting to render 3DGS remotely and dispatching it using uninterrupted, conventional video protocols~\cite{LLL-CAdViSE, shi2015survey} mitigates client rasterization requirements but inherently depends on temporal buffering and sequential inter-frame coding. In the context of unpredictable 6DoF navigation, these segment-oriented buffers yield unacceptable motion-to-photon (MTP) latencies, critically undercutting the user's interactive immersion. 

In this paper, we propose \textbf{TIGAS} (Thin-client Interactive Gaussian Adaptive Streaming), a testbed and remote rendering framework for interactive 3DGS streaming. TIGAS is explicitly designed to serve as the structural backbone for 3DGS streaming in scenarios where edge clients lack the computational power or network bandwidth requisite to render 3DGS autonomously. Leveraging a modular architecture, TIGAS allows plug in, isolation, and evaluation of arbitrary 3DGS models, ABR algorithms, user movement traces, network conditions, and neural upscaling/restoration models. Moreover, by offloading geometry-heavy processing and decoupling visual quality from client-side hardware, TIGAS effectively circumvents the tradeoff between long startup delays (downloading the full model) and quality degradation (downloading partial models). 

Instead of streaming geometry, TIGAS offloads rasterization to a GPU backend. The backend generates view-dependent 2D projections based on the client's 6DoF pose and streams them as images. To ensure a smooth interactive experience, TIGAS focuses on minimizing motion-to-photon (MTP) latency in remote rendering, a critical challenge for 6DoF interaction. First, it leverages QUIC~\cite{rfc9114alpn} as the transport protocol. Unlike TCP, which is often used in traditional DASH streaming via HTTP/1.1~\cite{rfc2616} or HTTP/2~\cite{rfc7540}, QUIC~\cite{rfc9000} mitigates Head-of-Line (HoL) blocking, ensuring that a lost packet pertaining to one frame does not stall the delivery of subsequent frames~\cite{yan2025instant}. Second, to adapt to fluctuating network conditions without inflating latency, TIGAS introduces a dedicated client-side, throughput-based Adaptive Bitrate (ABR) algorithm. Rather than buffering future frames, TIGAS's ABR dynamically adjusts the rendering resolution and compression quality to match the estimated network throughput, targeting a strict interaction latency threshold of \qty{100}{\milli\second}. 

An exploratory WebGPU Super Resolution (SR) pipeline is also included to study maximizing perceptual quality on the client. Experimental results using the dataset at~\cite{Kerbl2023} demonstrate that TIGAS enables real-time, interactive streaming of 3DGS models in real network conditions, at an average quality of 0.88 SSIM. 

In summary, the main contributions of this paper are that TIGAS combines: \textit{(i)} a GPU backend with multi-model and multi-user 3DGS management, optimized with persistent tensor residency, \textit{(ii)} a stateless per-frame JPEG request-response pipeline over QUIC, \textit{(iii)} a new Latency ABR which outperforms State of the Art (SOTA) controllers, and \textit{(iv)} a smart browser client that can apply WebGPU super-resolution/restoration before display.

\section{Related Work}
The rapid adoption of 3DGS has encouraged researchers to optimize its storage, transmission, and rendering efficiency. We categorize existing approaches into model compression, native streaming, and remote rendering.

\subsection{3DGS Compression and Simplification}
A 3DGS scene represents the environment as anisotropic Gaussian primitives, each defined by geometric attributes (center, scale, rotation) and appearance features (opacity, Spherical Harmonics) modeling view-dependent color and specular effects~\cite{Kerbl2023}. To reduce the memory footprint, recent work has applied quantization and pruning techniques to 3DGS primitives. \cite{fan2023lightgaussian} proposed to prune Gaussians based on their contribution to the final image, significantly reducing model size without major visual degradation. Similarly, \cite{girish2023eagles} and \cite{navaneet2024compgs} explored vector quantization and entropy coding to compress Gaussian attributes. While these methods reduce storage requirements, they do not alleviate the rendering bottleneck on the client device; the client must still decompress and rasterize the full set of primitives.

\subsection{Native 3DGS Streaming}
Native streaming architectures transmit the 3D model data to the client, often employing Level of Detail (LoD) or spatial partitioning to manage bandwidth. \cite{sun20253dgsabr} introduced LTS, a DASH-based framework that organizes Gaussians into layers (quality), tiles (space), and segments (time). LTS enables clients to request only the necessary subset of data based on their viewport. Similarly, LapisGS~\cite{shi2025lapisgs} utilized a layered approach with progressive rendering, allowing clients to download a coarse base layer followed by refinement layers. Certain frameworks, like GenStream~\cite{artioli2025genstream} attempt to resolve the 3DGS bandwidth bottleneck by pre-streaming the static environment model and transmitting only the essential attributes of dynamic scene actors in real-time, to be overlaid on the 3DGS model. Other works focus on the integration of 3DGS with standard video codecs. \cite{liu2024swings} and \cite{hu20254dgc} proposed encoding dynamic Gaussian attributes into video frames to leverage hardware video decoders. \cite{zhang2025streaminggsvoxelbasedstreaming3d} explored voxel-based partitioning to enable view-frustum culling during transmission. Despite these optimizations, native streaming inherently relies on the client's GPU for rasterization. As noted by \cite{Wei2025}, achieving high frame rates on edge devices often requires hardware-algorithm co-design, which is not feasible for generic web clients.

\subsection{Interactive Media Remote Rendering}
Remote rendering shifts the computational burden to the backend, streaming rendered 2D images or videos to the client. This approach is well-established in Cloud Gaming and Virtual Desktop Infrastructure (VDI)~\cite{shi2015survey}. However, standard video streaming protocols such as HLS/DASH are often ill-suited for the ultra-low latency required by 6DoF interaction~\cite{LLL-CAdViSE}. Furthermore, while Cloud Gaming platforms routinely achieve low-latency 6DoF streaming using WebRTC and hardware-accelerated video scaling, these represent locked-in, proprietary commercial applications rather than open research environments. More importantly, WebRTC's continuous statefulness struggles to support the extreme per-frame ABR flexibility required by erratic 6DoF Gaussian navigation~\cite{sun20253dgsabr}. Thus, recent architectures turn to stateless QUIC~\cite{rfc9000} to: \textit{(i)} eliminate HoL blocking for frequently synchronizing 6DoF updates natively, and \textit{(ii)} adopt built-in congestion control and rapid connection establishment (0-RTT) for fast image responses compared to traditional HTTP/1.1 or HTTP/2 stacks. 

While some modern remote rendering systems implement QUIC to carry traditional video streams, they remain bottlenecked by the inherent encoding and decoding latencies of inter-frame video codecs during rapid viewport changes. In contrast, TIGAS operates as an open-source testbed that shifts the computational burden entirely to a remote GPU backend utilizing a stateless, per-frame JPEG pipeline over QUIC. While this approach deliberately trades bandwidth efficiency for absolute zero head-of-line blocking and instant ABR profile switching, eliminating the need to wait for IDR/I-frames and avoiding encoder statefulness, it perfectly aligns with the strict sub-\qty{100}{\milli\second} motion-to-photon latency requirements of 6DoF interactivity. By introducing an ABR algorithm that enforces this threshold on a frame-by-frame basis, TIGAS completely decouples visual quality from the hardware constraints of client devices. This shift to adaptive, view-dependent stateless 2D projection streaming provides a robust solution for responsive 6DoF navigation on commodity hardware when compared to segmented video transport.

\begin{figure}[t]
    \centering
    \includegraphics[width=\linewidth]{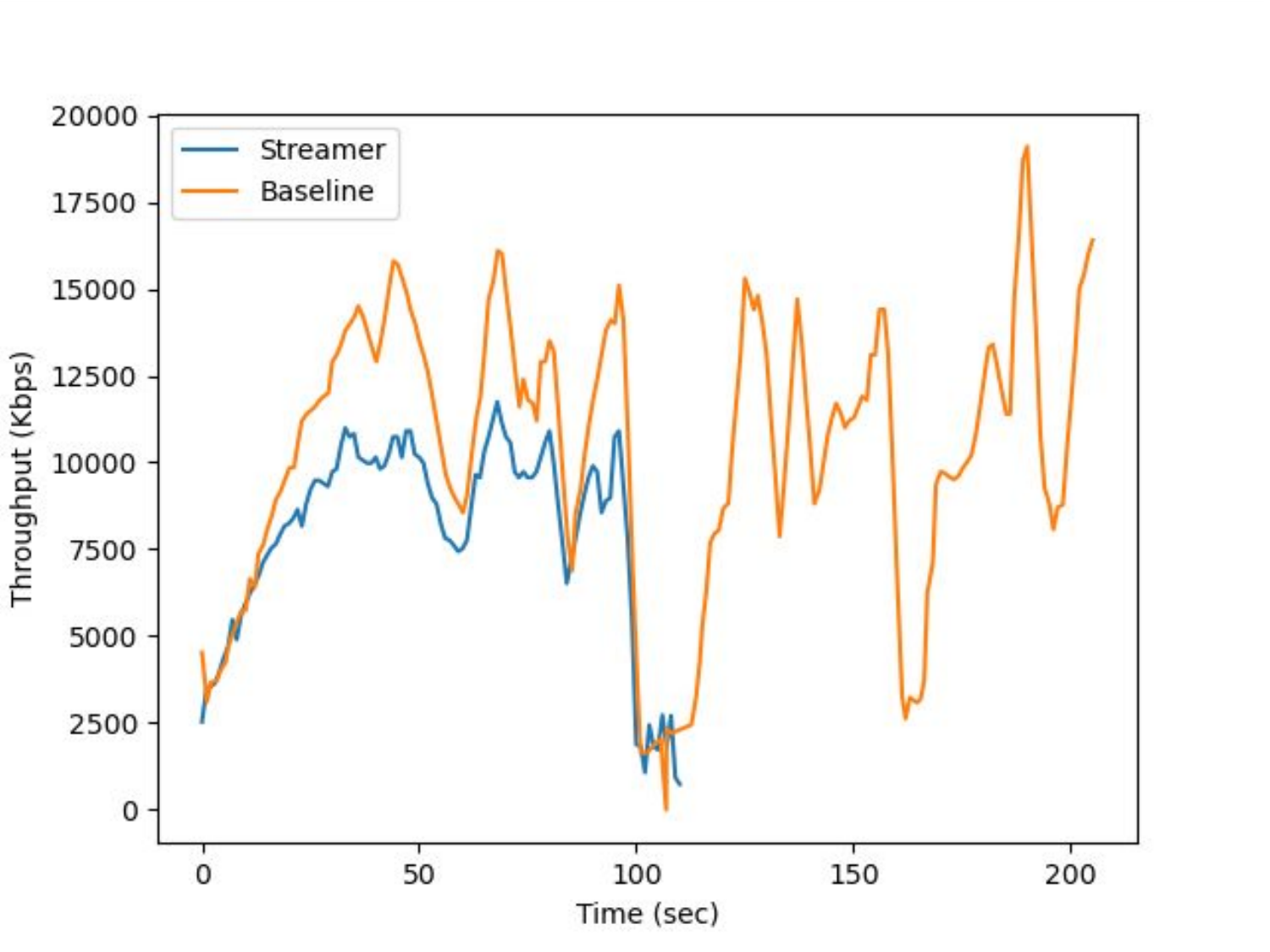}
    \caption{Network bandwidth utilization over time during a simulated navigation session using an EyeNavGS trace. While the baseline monopolizes bandwidth for an extended period, TIGAS completes the interaction session using only a fraction of the capacity.}
    \label{fig:progressive_throughput}
\end{figure}

\section{Motivating Example}
To illustrate the necessity of an adaptive remote rendering framework for 3DGS, we compare TIGAS against two baselines: \emph{(i)} progressive download for local rendering, and \emph{(ii)} traditional video streaming (DASH/CMAF). In progressive downloads, clients must download massive geometry prior to rendering. Figure~\ref{fig:progressive_throughput} underscores this inefficiency: while the baseline saturates capacity endlessly just to acquire the model, TIGAS operates comfortably below total capacity. For an interaction driven by a real EyeNavGS trace~\cite{ding2025eyenavgs}, TIGAS streams only the requested projections and finishes before the baseline downloads half of the data. Conversely, treating remote 3DGS rendering as standard video streaming (e.g., a low-latency DASH/CMAF baseline) reduces bandwidth via inter-frame compression but introduces segment playback latency spiking above \qty{500}{\milli\second}, critically failing the sub-\qty{100}{\milli\second} 6DoF requirement. This necessitates a stateless, decoupled architecture like TIGAS.

\begin{figure*}[t]
    \centering
    \includegraphics[width=\textwidth]{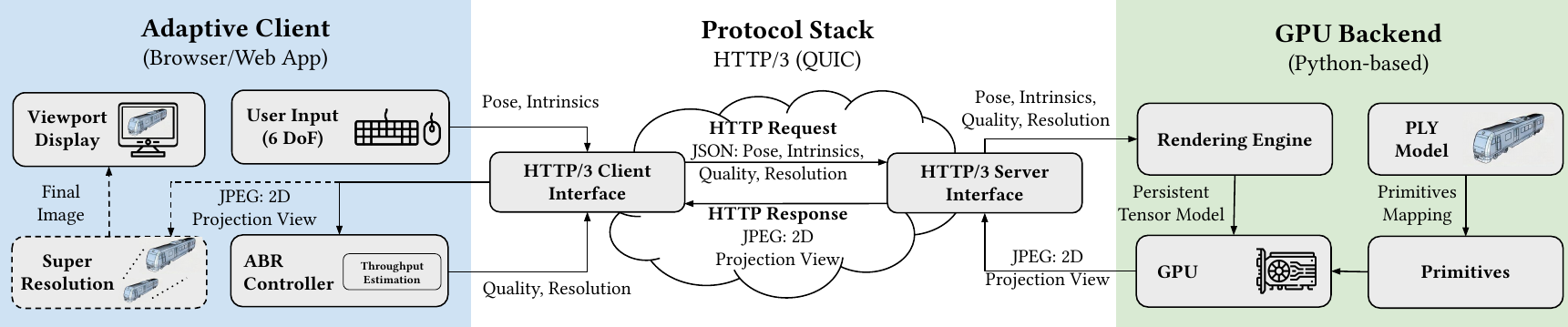}
    \caption{TIGAS consists of an Adaptive Client that handles 6DoF user input and ABR logic, connected via HTTP/3 (QUIC) to a GPU backend performing rasterization, thus offloading rendering from the client.}
    \label{fig:system_architecture}
\end{figure*}

\section{TIGAS System Design}
Figure~\ref{fig:system_architecture} shows the architecture of TIGAS comprising three logical components: \textit{(i)} a \textit{GPU Backend} with a \textit{Rendering Engine}, \textit{(ii)} a low-latency \textit{HTTP/3 (QUIC) Protocol Stack} for transmission, and \textit{(iii)} an \textit{Adaptive Client}. 

\subsection{GPU Backend}
\subsubsection*{PLY Model and Primitives.}
TIGAS manages a directory of 3DGS models. At startup, the backend scans available model folders, registers metadata (identifier, name, preview image), and lazily loads model tensors on demand, \ie only when a client requests one. Each PLY scene is parsed into 3D Gaussian primitives (means, quaternions, scales, opacity, spherical harmonic (SH) coefficients) and transferred to GPU memory for rendering. To keep memory usage bounded across many scenes, TIGAS includes an eviction loop that unloads inactive models after an inactivity timeout. 

\subsubsection*{Rendering Engine.}
Synthesizing 2D viewpoints from 3DGS scenes under real-time 6DoF interacton relies on a tile-based rasterizer projecting primitives onto a 2D image plane. The engine supports rendering ABR-driven, downscaled, and resampled representations on the fly to control transmission payload sizes. 

\subsubsection*{GPU Backend.} 
To eliminate recurrent I/O latency bottlenecks, the system prioritizes persistent VRAM residency for initialized Gaussian primitives, pre-loading them prior to interaction. 

\subsection{Protocol Stack}
The primary transport employs per-frame stateless requests over QUIC. The client continuously sends compact JSON payloads carrying pose, camera intrinsics, and its desired ABR target profile determined from a predefined bitrate ladder of resolution and JPEG quality pairs. The backend serves each request as an isolated rendering task bounding frame queue buildup and limits HoL effects. This ensures rapid 6DoF camera updates are immediately reflected in the incoming network flight without stalling on previous historical data. 

\subsection{Adaptive Client} 
The browser-native client is designed for commodity devices, supporting standard frame decoding, swappable WebGPU super-resolution models based on NEVES~\cite{neves}, and modular ABR algorithms. 

\subsubsection*{User Input.} 
The client decouples the rendering loop from the input loop by continuously polling user input for pose requests. This decoupling minimizes MTP latency, ensuring that visual feedback aligns tightly with user keystrokes for an engaging parsing experience irrespective of network jitter. 

\subsubsection*{ABR Controller.} 
To handle network fluctuations, the client integrates an ABR algorithm that selects the optimal rendering resolution and compression per frame. This ensures the total request time remains comfortably within the \qty{100}{\milli\second} latency threshold typical for instantaneous human perception~\cite{8743342}. 

\section{Implementation}
We implement TIGAS as a client-backend system consisting of a Python-based rendering service and a browser-based frontend. 

\subsection{GPU Backend}
\subsubsection*{PLY Model and Primitives.} Each scene is parsed from PLY into pytorch~\cite{ketkar2021pytorch} tensors mapped to the compute device. For each Gaussian $n$, TIGAS processes attributes as follows:
\begin{itemize}[leftmargin=*,nosep]
    \item \textbf{Location \& Scale.} The centers $\boldsymbol{\mu}_n \in \mathbb{R}^3$ are stacked into a tensor $\mathbf{M} \in \mathbb{R}^{N \times 3}$. Scales $\mathbf{s}_n$ are stored as log-vectors and exponentiated element-wise $\exp(\mathbf{s}_n)$ to enforce positive 3D dimensions.
    \item \textbf{Rotation \& Opacity.} Orientation relies on unit quaternions $\mathbf{q}_n \in \mathbb{R}^4$. Opacities in logit space transform via a sigmoid function $\sigma(x)$ to the $[0, 1]$ interval.
    \item \textbf{Color and SH.} SH coefficients form a tensor $\mathbf{C} \in \mathbb{R}^{N \times 16 \times 3}$. We default to degree $d_{\text{SH}} = 0$ (view-independent color) to minimize baseline bandwidth, though up to degree 3 is supported.
\end{itemize}

\subsubsection*{Rendering Engine.} 
To synthesize views, world-space Gaussian centers must project onto a 2D camera plane. The client transmits \emph{azimuth} $\theta$ and \emph{elevation} $\phi$ to construct a rotation matrix $\mathbf{R} \in \mathbb{R}^{3\times 3}$. The world-to-camera transformation matrix $\mathbf{T}_{w \rightarrow c}$ is defined as:
\[\mathbf{T}_{w \rightarrow c} = \begin{bmatrix}\mathbf{R}^\top & -\mathbf{R}^\top\mathbf{t}\\ \mathbf{0}^\top & 1\end{bmatrix} \in \mathbb{R}^{4\times 4},\]
mapping world centers to camera coordinates $\mathbf{p}_c = [x_c, y_c, z_c]^\top$. The points map to pixel coordinates $(u, v)$ via an intrinsics matrix $\mathbf{K}$ governed by focal lengths and principal points:
\[\lambda \begin{bmatrix} u & v & 1 \end{bmatrix}^\top = \mathbf{K} \mathbf{p}_c, \quad \text{where } \lambda = z_c.\]
TIGAS dynamically re-centers principal points $c_x, c_y$ when the ABR modifies the frame resolution limit. Final rasterized frames are transferred as 8-bit JPEG payloads over HTTP/3 or pushed iteratively to \texttt{ffmpeg} in our DASH chunking baseline without relying on browser cache storage. 

\subsubsection*{Dynamic Multi-Model Runtime.}
The backend exposes model discovery and explicit loading endpoints. A model is loaded on first access, reference-counted during rendering, and periodically considered for eviction when inactive. This sharing mechanism enables interactive switching across many scenes without preloading all models into VRAM. Crucially, it inherently supports a multi-client architecture: any number of incoming users can simultaneously request renders from a shared model already loaded into GPU memory, eliminating redundant VRAM overhead and bypassing repeated loading bottlenecks for concurrent interactive sessions. While GPU memory utilization scales gracefully due to shared model tensors, the required concurrent rasterization and encoding compute scales linearly per client; we note that investigating horizontal scaling methodologies across multi-GPU pools represents a fundamental step toward massive production environments. 

\subsection{Protocol Stack: HTTP/3 over QUIC}
TIGAS operates an ASGI application backed by \texttt{aioquic} with HTTP/3 and \texttt{reno} congestion control~\cite{rfc9114alpn,rfc5681congestion}. While currently using reliable HTTP/3 streams for request-response pairs, it supports QUIC datagrams (up to \qty{64}{KB}) to drop delayed frames upon packet loss, similar to RTP. QUIC events map natively to ASGI scopes, feeding asynchronous tasks into thread executors to avoid loop-blocking during GPU interactions. In the stateless regime, the backend issues single, timestamped JPEG frames per client camera-pose hit. Concurrently, a separate DASH integration broadcasts fragmented frames toward a CMAF packager. 

\begin{figure*}[t]
    \centering
    \begin{subfigure}[t]{0.31\linewidth}
        \centering
        \includegraphics[width=\linewidth]{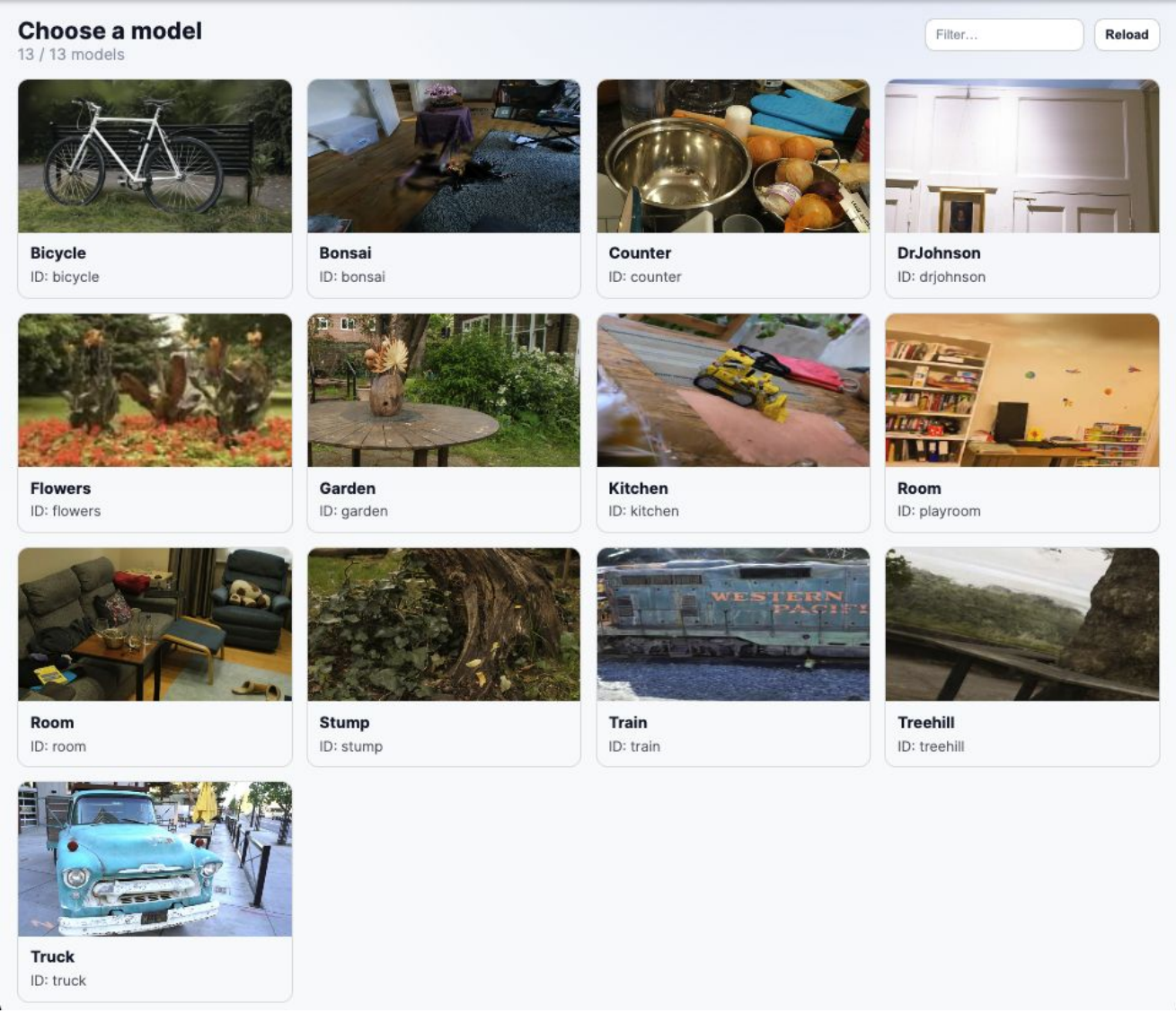}
    \end{subfigure}\hfill
    \begin{subfigure}[t]{0.36\linewidth}
        \centering
        \includegraphics[width=\linewidth]{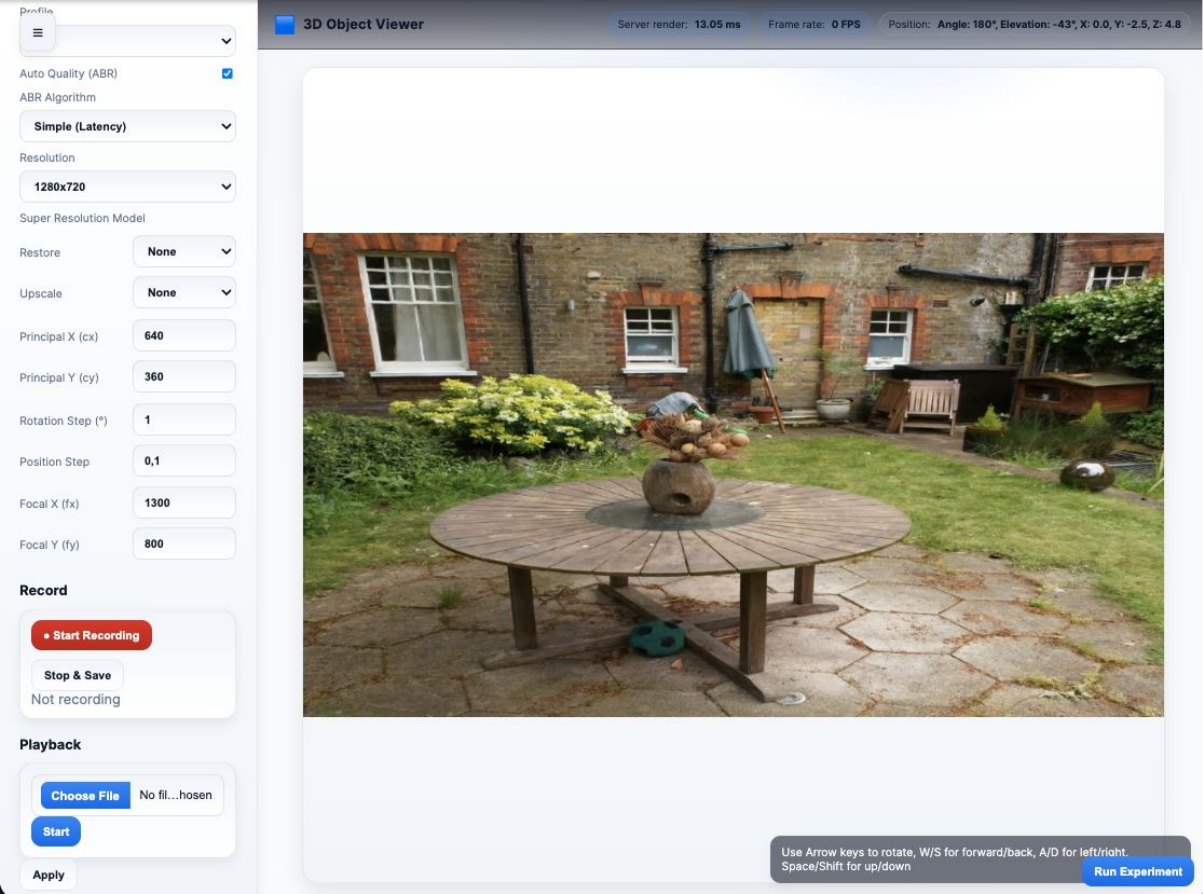}
    \end{subfigure}\hfill
    \begin{subfigure}[t]{0.32\linewidth}
        \centering
        \includegraphics[width=\linewidth]{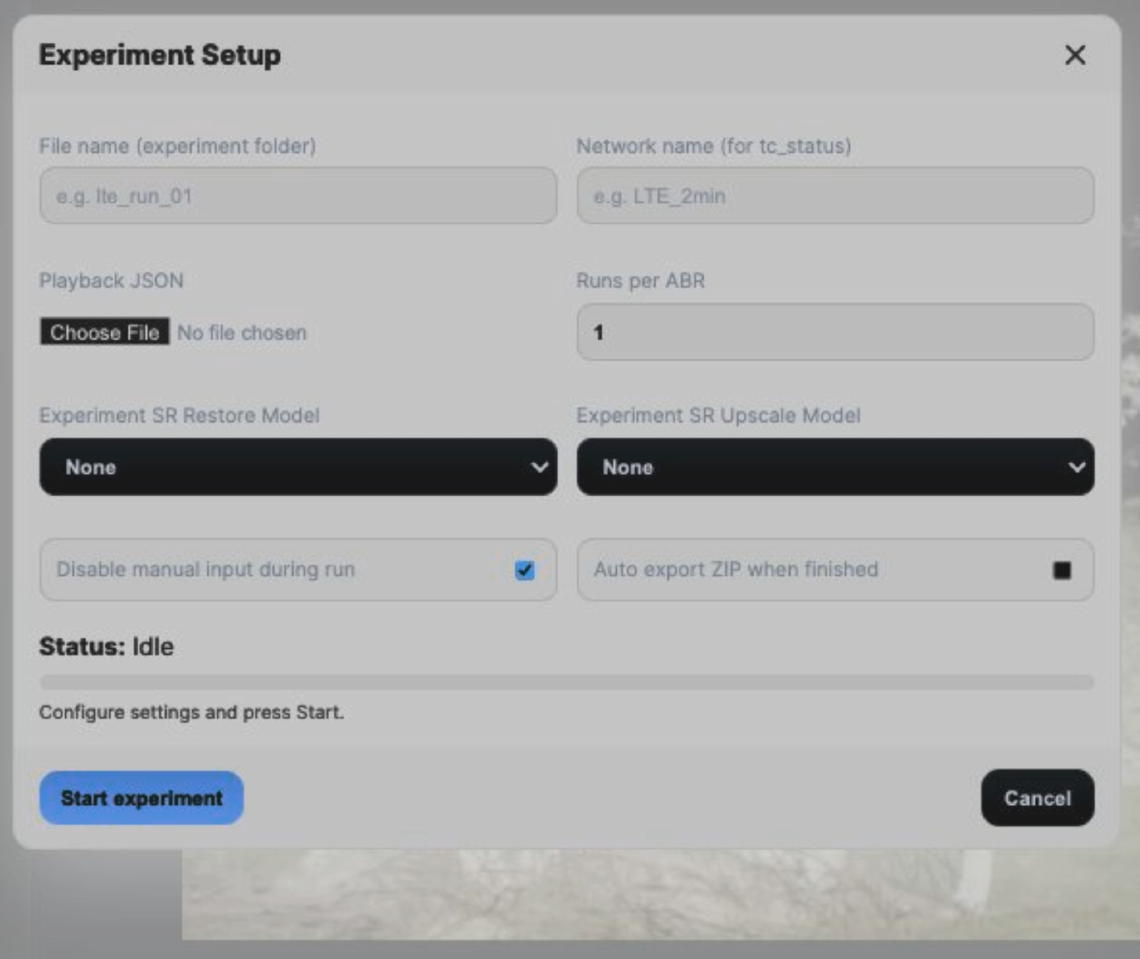}
    \end{subfigure}
    \caption{Screenshots of the client web browser interface: (a) model selection screen, (b) main client UI, and (c) automated experiment setup popup.}
    \label{fig:client_ui}
\end{figure*}

\begin{figure*}[tbh]
    \centering
    \includegraphics[width=\linewidth]{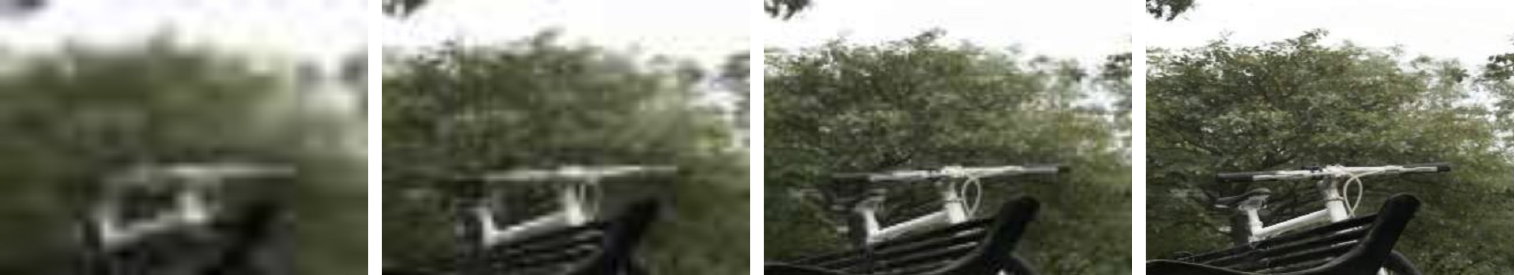}
    \caption{Visual comparison of rendered frames across the different quality profiles defined in Table~\ref{tab:qualities}. From left to right: Low (profile 3), Medium-Low (profile 2), Medium-High (profile 1), High (profile 0).}
    \label{fig:quality_comparison}
\end{figure*}

\subsubsection*{Client Interface.} 
The client-side implements a small web application using the ASGI framework Starlette~\cite{starlette_framework}. The app mounts: (i) a GET route at \texttt{/} which serves the single-page (HTML) client, (ii) a POST route at \texttt{/render} which returns a freshly rendered JPEG, and (iii) a static files mount at \texttt{/static} for JS/CSS and the 3DGS model. The client issues \texttt{POST /render} with a JSON body containing the camera pose (azimuth $\theta$, elevation $\phi$, and translation $\mathbf{t} = [t_x, t_y, t_z]^\top$), intrinsics (focal lengths $f_x, f_y$, and principal point $c_x, c_y$), and target image resolution ($W, H$). 

\subsection{Adaptive Client}
\subsubsection*{Model Selection.}
To begin a streaming session, users interact with the model selection dashboard, illustrated in Figure~\ref{fig:client_ui}(a). This menu displays visual thumbnails of all available 3DGS scenes hosted on the backend, allowing users and researchers to rapidly load or swap between scenes. 

\subsubsection*{Viewport Display.}
The main HTML client, shown in Figure~\ref{fig:client_ui}(b), offers a collapsible sidebar with fields for $f_x, f_y, c_x, c_y$, resolution, and step sizes. On each update, the client assembles JSON with pose, intrinsics, and resolution, then \texttt{fetch}es \texttt{/render}. The response body, i.e., a JPEG image, is then in the viewport. Eventual errors replace the viewport with a short message; URLs are revoked after use to avoid leaks. The focal length and initial field-of-view (FoV) are established at the beginning of the session. During ABR adaptations, the system adjusts the rendering resolution and JPEG compression while proportionally scaling the camera intrinsics ($f_x, f_y, c_x, c_y$) to guarantee that the viewport and FoV remain constant, preventing any perceptual popping or misalignment. 

\subsubsection*{Experiment Automation.}
To facilitate reproducible evaluation, the web application includes an automated experiment setup interface, depicted in Figure~\ref{fig:client_ui}(c). This popup allows researchers to configure scheduled experimental runs, inject real 6DoF movement traces, and manage automated metric gathering sessions. 

\subsubsection*{User Input.} 
Keyboard controls update pose: arrows adjust azimuth/elevation; \texttt{W/S} move forward/backward; \texttt{A/D} move sideways; \texttt{Shift/Space} move up/down. Motion in the $x$-$z$ plane is computed relative to azimuth $\theta$ using $\sin\theta$ and $\cos\theta$ so ``forward'' follows the camera's current facing. An animation loop polls keys and triggers updates without blocking the UI. 

\subsubsection*{Throughput Monitoring.} 
For each request $i$, the client records start/end times and reads the \texttt{Content-Length} header to estimate the image size $S_i$ in bytes. Instantaneous throughput is computed as $\hat{r}_i = \frac{S_i}{\Delta t_i}$, where $\Delta t_i$ is the download duration in seconds for request $i$. To reduce jitter, TIGAS maintains an Exponential Moving Average (EMA) over the last few samples (default history length is $5$ requests) with a default smoothing factor $\alpha = 0.3$. 

\begin{table}[pb]
    \centering
    \caption{Frame Quality Levels (Bitrate Ladder)}
    \begin{tabular}{lccc}
        \toprule
        Quality (Profile) & Resolution & JPEG Comp. (\%) & Size (KB) \\
        \midrule
        Low (3)         & $320 \times 180$ & 10\% & 6--8   \\
        Medium--Low (2) & $640 \times 360$ & 35\% & 18--22   \\
        Medium--High (1)& $960 \times 540$ & 65\% & 50--60 \\
        High (0)        & $1280 \times 720$ & 90\% & 180--300 \\
        \bottomrule
    \end{tabular}
    \label{tab:qualities}
\end{table}

\begin{figure}[t]
    \centering
    \includegraphics[width=\linewidth]{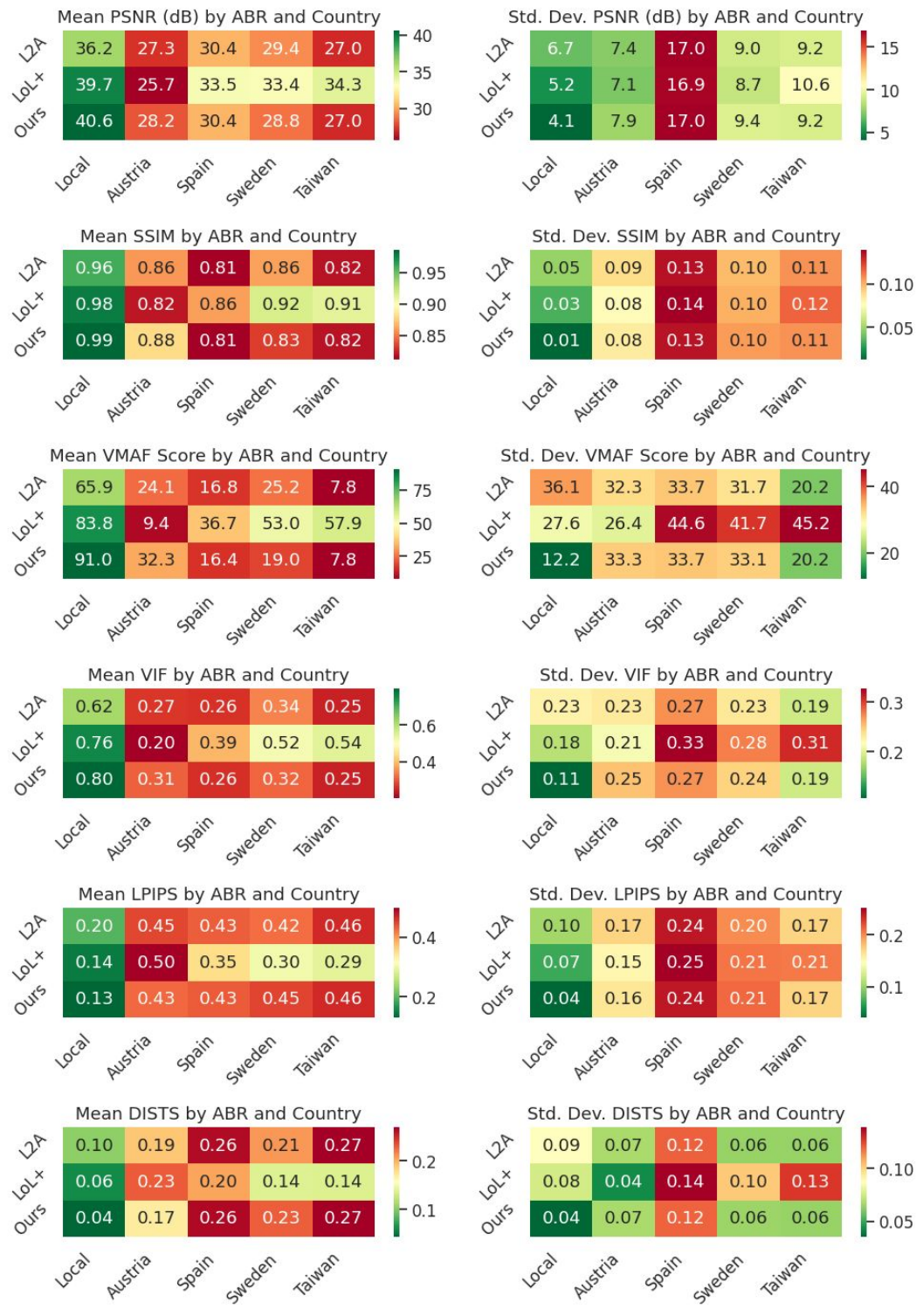}
    \caption{Average quality per metric across ABR algorithms and client locations.}
    \label{fig:quality_heatmap_original}
\end{figure}

\begin{figure}[t]
    \centering
    \includegraphics[width=\linewidth]{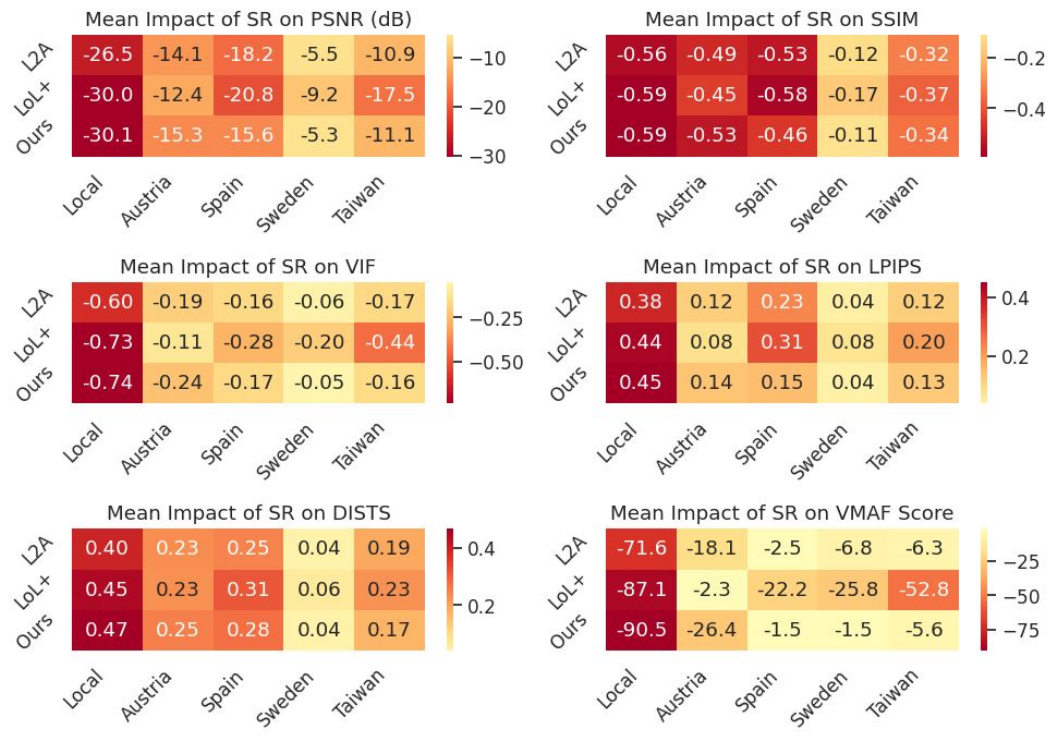}
    \caption{Impact of super resolution on frame quality compared to original non-SR frames.}
    \label{fig:quality_heatmap_sr_difference}
\end{figure}

\begin{figure}[t]
    \centering
    \includegraphics[width=\linewidth]{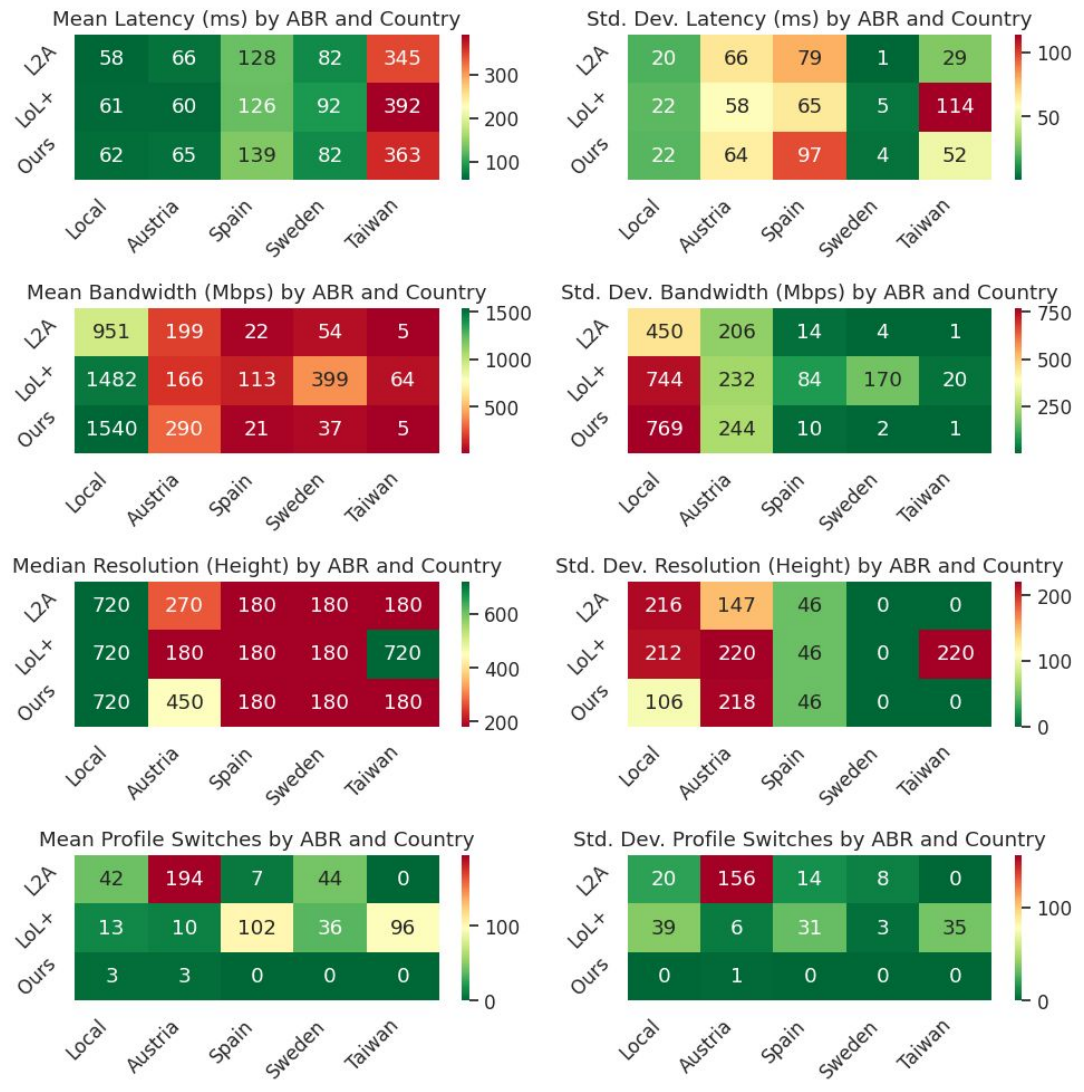}
    \caption{System performance metrics comprising latency, rendering time, resolution, bandwidth, and profile switching dynamics.}
    \label{fig:quality_heatmap_others}
\end{figure}

\subsubsection*{ABR Controller.} 
To manage network fluctuations without interrupting the 6DoF experience, we propose a receiver-driven ABR algorithm, hereafter referred to as the \textit{Latency} ABR. Our algorithm utilizes a predefined bitrate ladder (see Table~\ref{tab:qualities}) that maps each quality level $\ell$ to a specific JPEG compression level and its corresponding expected size $\tilde{S}_\ell$. For any candidate level, the estimated completion time $\tilde{T}_\ell$ (in seconds) is calculated based on the current throughput $r_i$:
\[\tilde{T}_\ell = \frac{\tilde{S}_\ell}{r_i \cdot 1024}\]
The \textit{Latency} algorithm uses this estimation to target a maximum per-request time of $T_{\text{target}} = 100$~ms, which represents the threshold for instantaneous human perception. To maintain stability during 6DoF navigation, the ABR policy adheres to the following mathematical constraints:
\begin{itemize}[leftmargin=*,nosep]
    \item \textit{Upgrade Condition:} A switch to a higher resolution level $\ell+1$ occurs only if the predicted time satisfies $\tilde{T}_{\ell+1} \leq T_{\text{target}}$.
    \item \textit{Downgrade Condition:} To mitigate the impact of network jitter, a downgrade to level $\ell-1$ is triggered if the current completion time consistently exceeds the safety margin, defined as $\tilde{T}_\ell > T_{\text{margin}}$.
    \item \textit{Safety Constraint:} We maintain the relationship $T_{\text{target}} < T_{\text{margin}}$ to create a deadband that prevents frequent oscillations.
\end{itemize}
To enhance stability, a hold counter requires $h = 3$ consecutive requests before finalizing a resolution change. This is bypassed during rapid panning to allow immediate downgrades. When ABR is active, the manual resolution UI is disabled to ensure consistent performance. Following any resolution switch, TIGAS automatically scales the principal point and focal lengths to the new resolution, keeping the 2D projection centered and the FoV constant. 

\subsubsection*{Smart Thin Client with WebGPU.}
To decouple perceptual quality from network payload size, TIGAS optionally applies browser-side neural restoration and upscaling after frame reception. As shown in Fig~\ref{fig:system_architecture} with a dashed line style, the client integrates WebGPU models adapted from the NeVES enhancement pipelines~\cite{neves} and can chain restore and upscale stages. This enables low-bandwidth frames to remain visually acceptable. While client-side SR makes the client somewhat ``less thin'', SR is a heavily optimized task for modern Neural Processing Units (NPUs) found in commodity devices, compared to the unstructured compute demands of 3DGS rasterization~\cite{liu2025duplexgs, rai2025uvgs}. Currently, however, the freely available generalized NPU shaders are not tailored to exact 3DGS point artifacts and can induce temporal flickering. Thus, we primarily architect WebGPU integration as a forward-looking capability: as sophisticated 3DGS-specific upsampling algorithms become accessible, they can be immediately hot-swapped into TIGAS's zero-delay client application. Besides, SR is entirely optional, and can be activated or dismissed dynamically based on underlying device power limits without fracturing the baseline remote rendering pipeline. 

\subsubsection*{Sampled OPFS Capture and Post-Session Materialization.}
To avoid runtime interference, TIGAS samples enhanced client frames at a fixed stride and stores them in OPFS via a worker thread. After playback, sampled SR frames are uploaded to the backend. The backend then replays logged movements at the same sampled indices and materializes: \textit{(i)} original transmitted JPEG and \textit{(ii)} deterministic ground-truth PNG. The triplet (original, SR, GT) is used for offline quality evaluation. Specifically for VMAF, the captured independent frames are mapped back into sequential uncompressed video containers to apply the established temporal and spatial algorithms against the baseline. 

\section{Performance Evaluation}
\subsubsection*{Setup.}
To demonstrate TIGAS's capabilities as a comprehensive and scalable testbed, we evaluate the system using two distinct experimental environments to isolate network effects and assess real-world performance. In the \textit{local setup}, both the backend and client run on a single machine. We utilize two such local machines equipped with an NVIDIA Quadro T1000 and an NVIDIA RTX 2060 Super GPU, respectively. This configuration eliminates uncontrolled network fluctuations by enforcing a predetermined network trace (the \texttt{bus0001} trace from~\cite{vanderHooft2016nettraces}) utilizing the Linux \texttt{tc} command, enabling highly controlled benchmarking. 

In the \textit{distributed setup}, we deploy a remote backend in Austria, equipped with an NVIDIA Quadro RTX 8000 and connect client machines located in Austria (Apple M1 Pro), Spain (Apple M4), Sweden (NVIDIA RTX 4050), and Taiwan (two clients equipped with NVIDIA RTX 3050 and RTX 2080 Ti, respectively). For the Taiwan experiments, the reported results represent the aggregate of the results across these two clients. This distributed configuration allows us to evaluate how TIGAS is impacted by latency as geographical distance increases under actual, unthrottled network conditions. Across both setups, we utilize 13 diverse 3DGS models made available by~\cite{Kerbl2023} and the bitrate ladder detailed in Table~\ref{tab:qualities}. To ensure realistic and reproducible benchmarking, we simulate client input using more than 100 real 6DoF movement trajectories collected from the EyeNavGS~\cite{ding2025eyenavgs} dataset. These benchmarking loops are fully orchestrated via the automated experiment setup interface detailed in Figure~\ref{fig:client_ui}(c). We compare TIGAS's Latency ABR against SOTA algorithms, namely LOL+~\cite{Lim2020lolplus} and L2A~\cite{Karagkioules2020L2A}. In total, we executed 27 local tests (9 per ABR) and 126 remote experiments, generating a total of $425{,}376$ rendered 3DGS views.

\subsection*{Results.} 
Figure~\ref{fig:quality_heatmap_original} summarizes TIGAS's baseline performance by showing the average quality across six distinct metrics (PSNR, SSIM, VMAF, VIF, LPIPS, and DISTS), categorized per ABR algorithm and per country. Note that these values reflect the original, non-super-resolved frames, thereby illustrating the raw image quality supplied by the remote backend based strictly on the ABR's capacity decisions. We include perceptual metrics like LPIPS, VIF, and DISTS to provide a deeper understanding of structural deformations and scaling artifacts, which are notoriously difficult for standard metrics like PSNR to capture accurately in novel 3D view synthesis. 

From this figure, we observe that the metrics mostly agree with each other. As expected, the local experiments achieve the highest baseline quality due to the stability of the simulated network trace compared to real-world, unthrottled live networks. For the distributed setup, one would expect the Austrian client to exhibit the best quality since it is geographically closest to the backend. However, the results are rather mixed: it records the lowest PSNR scores across all ABR algorithms, while securing comparatively strong scores in the other metrics (with the exception of LoL+). This behavior likely stems from real-network congestion and potential client-side processing bottlenecks when handling high volumes of UDP packets at very short ranges. Notably, LoL+ achieves substantially better performance in most metrics for clients located farthest from the backend, a counter-intuitive finding that necessitates deeper investigation in future work. 

Furthermore, analyzing the dataset's standard deviations reveals distinct variability profiles among the metrics. Taking Austria as an example, established pixel-based and feature-based metrics (PSNR, SSIM, LPIPS, and DISTS) present standard deviations generally proportional to their average values. In sharp contrast, variability is exceptionally high for VMAF and VIF. This is expected considering VMAF was designed inherently as an inter-frame video metric heavily reliant on temporal consistency. Because TIGAS serves independent JPEGs across fluctuating compression profiles, VMAF harshly penalizes this stateless spatial scaling and flickering loss of natural image information, proving highly sensitive to the temporal instability inherent in a stateless pipeline compared to traditional frame-by-frame structural metrics. 

In Figure~\ref{fig:quality_heatmap_sr_difference}, we gauge the impact of client-side super resolution (SR) on visual quality. Interestingly, applying current SR models generally results in a quality drop compared to the original frames. This decline stems from the computationally intensive burden on "thin" clients causing frame drops or abrupt flickering to maintain low latency, alongside a lack of model specialization, as current WebGPU SR shaders are trained on generic content rather than the intricate anomalies of 3DGS point-splatting. Nevertheless, we highlight this SR capability not as a purely experimental extra, but as a fully developed, integral feature of the TIGAS testbed. It demonstrates that the client architecture is ready to natively host and evaluate advanced, 3DGS-tailored WebGPU SR models. 

Figure~\ref{fig:quality_heatmap_others} illustrates detailed system and network metrics across four fundamental factors: 
\begin{itemize}[leftmargin=*,nosep]
    \item \textbf{(a) Latency.} As expected, interaction latency correlates predominantly with the geographical distance between the client and the backend, rather than the choice of ABR logic. Both the local and Austrian setups experience approximately \qty{60}{\milli\second} of total latency. We break down this \qty{60}{\milli\second} floor using empirical measurements and findings from the literature: the remote GPU rasterization takes about \qty{11}{\milli\second} and \qty{17}{\milli\second} on the RTX 2060 Super and Quadro T1000, respectively, and hardware-accelerated JPEG encoding takes under \qty{1}{\milli\second} on modern CUDA devices~\cite{Abouelhamayed2024jpegencodingspeed}. Overall latency remains successfully below the \qty{100}{\milli\second} interactivity threshold even for clients as far as Sweden. However, it surpasses this boundary for Spain and exceeds \qty{300}{\milli\second} for clients in Taiwan. 
    \item \textbf{(b) Average Bandwidth.} Average bandwidth consumption highlights the influence of transport distance. The local environment leverages Gbps-level capacity, while utilized bandwidth drops precipitously for restricted Taiwanese clients. While a \qty{5}{Mbps} connection comfortably supports standard chunked video streaming, TIGAS requires conditionally larger network allocations because it is computationally limited to stateless intra-frame JPEG compression. 
    \item \textbf{(c) Median Resolution.} Median requested resolutions precisely mirror the bandwidth patterns. The local configuration rapidly requests the highest available resolutions, closely followed by Austria. 
    \item \textbf{(d) Quality Switches.} Profiling switching behavior reveals the stability trade-offs. Although \textit{L2A} achieves comparable visual quality to peers, it suffers from excessive profile switches, deteriorating user QoE through structural popping. \textit{LOL+} is more stable but still applies an aggressive switching policy. Conversely, our Latency ABR actively mitigates jarring switches through strict temporal holds. 
\end{itemize}

\begin{figure}[t]
    \centering
    \begin{subfigure}[b]{0.49\linewidth}
        \includegraphics[width=\linewidth]{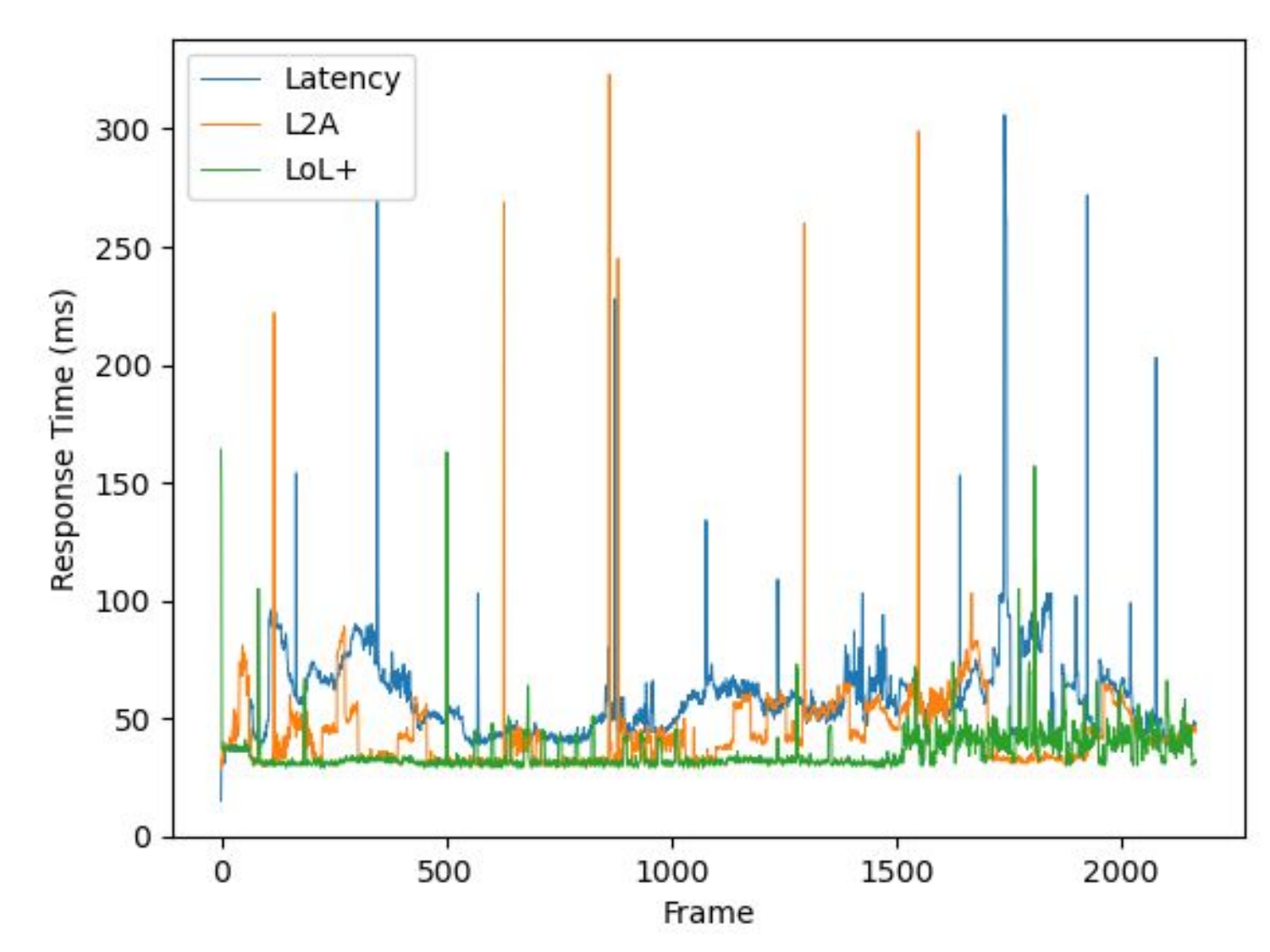}
        \caption{Per-frame response (\qty{20}{\milli\second} RTT).}
        \label{fig:response_time_20ms}
    \end{subfigure}
    \hfill
    \begin{subfigure}[b]{0.49\linewidth}
        \includegraphics[width=\linewidth]{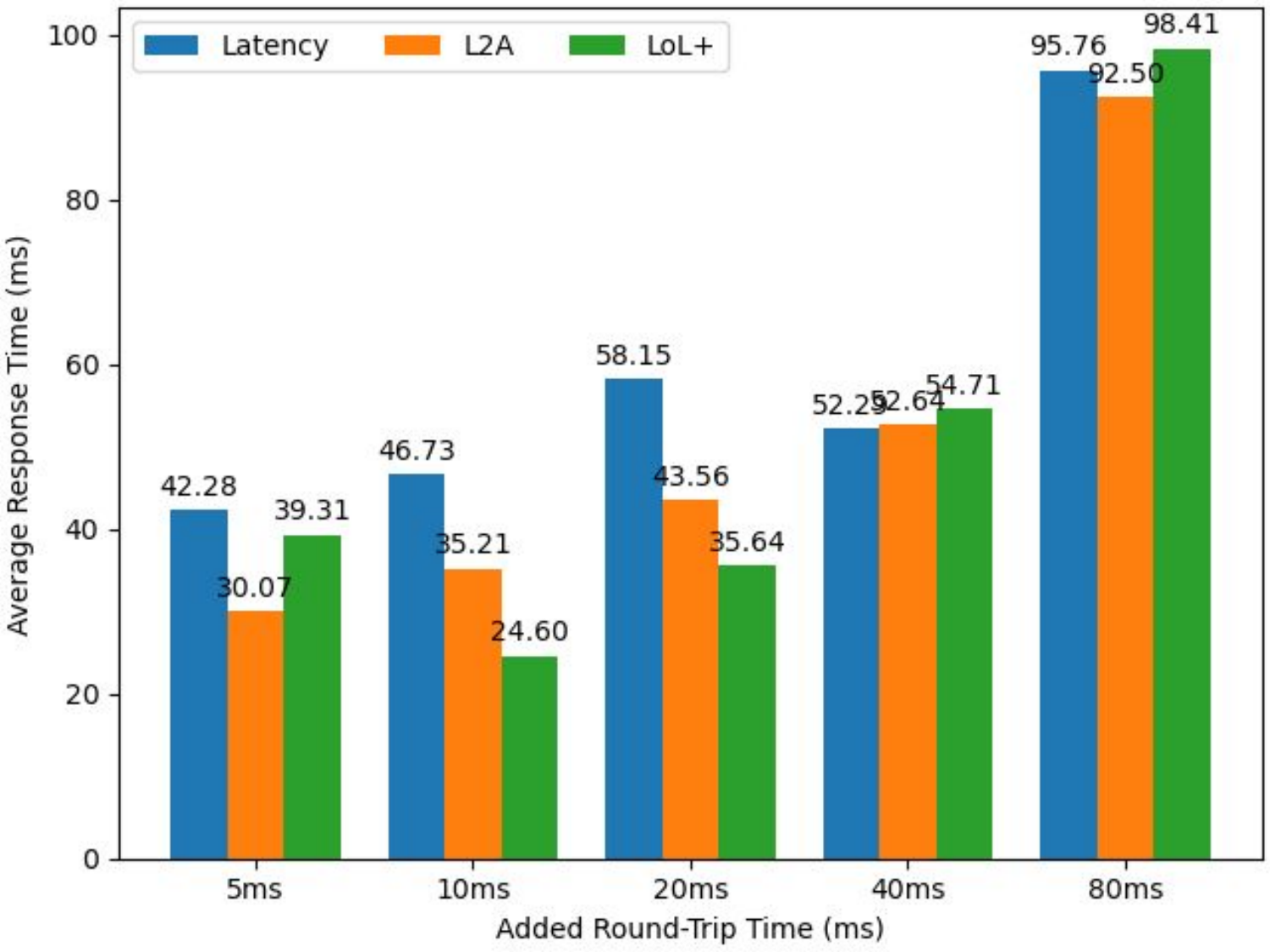}
        \caption{Avg response time across RTTs.}
        \label{fig:avg_response_time_rtt}
    \end{subfigure}
    \begin{subfigure}[b]{0.49\linewidth}
        \includegraphics[width=\linewidth]{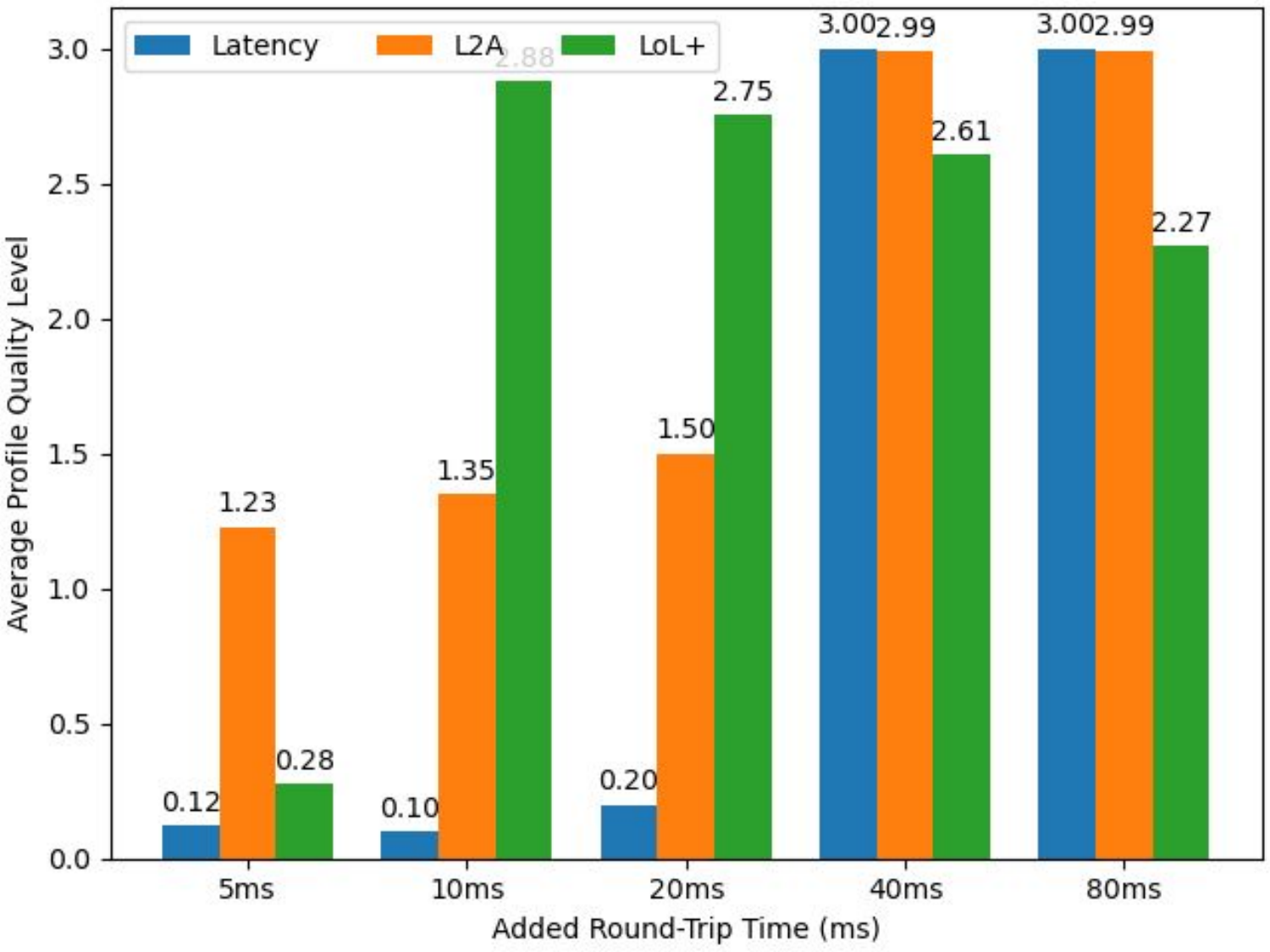}
        \caption{Avg profile across RTTs.}
        \label{fig:avg_profile_level}
    \end{subfigure}
    \hfill
    \begin{subfigure}[b]{0.49\linewidth}
        \includegraphics[width=\linewidth]{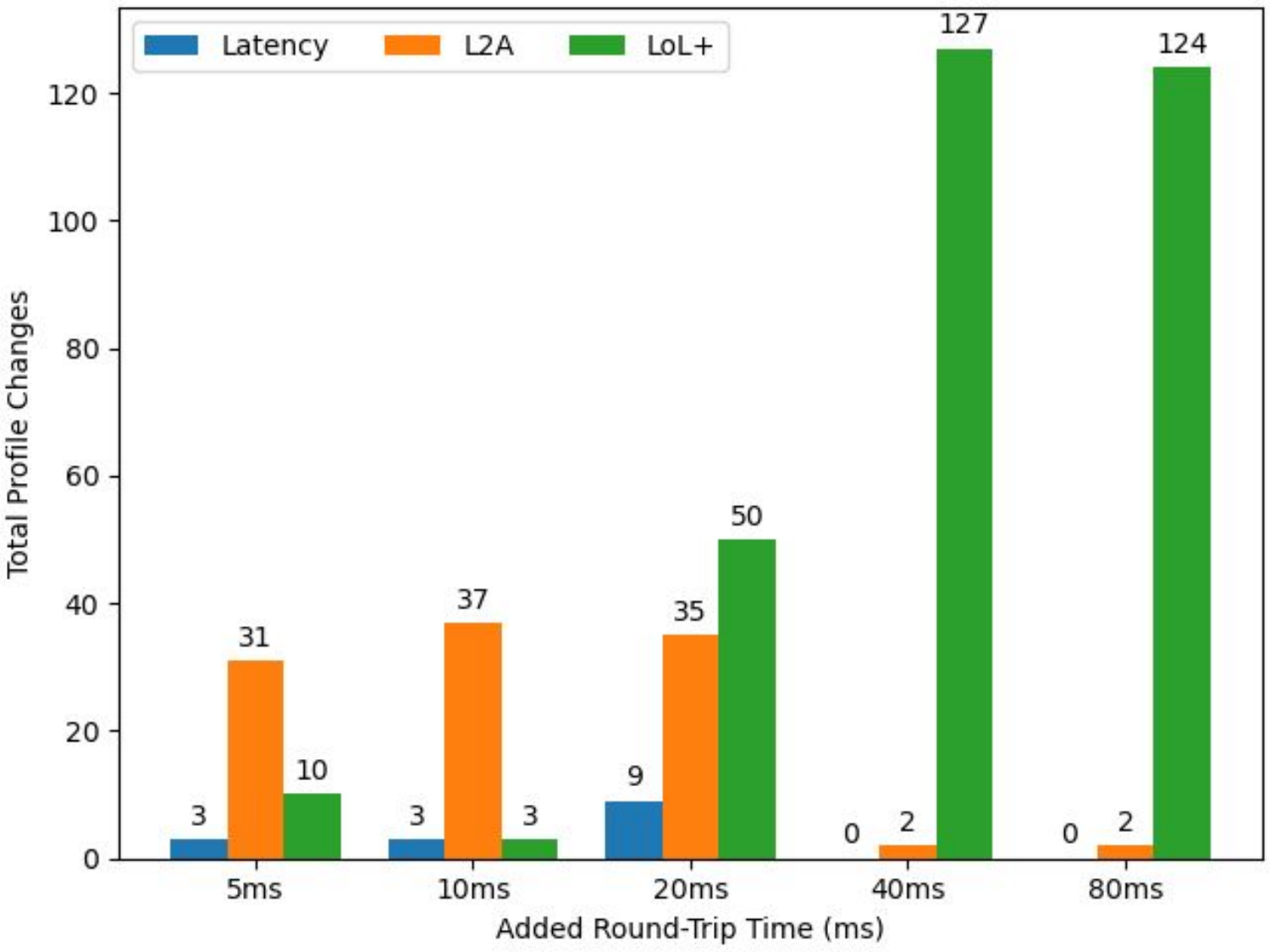}
        \caption{Profile changes across RTTs.}
        \label{fig:profile_changes}
    \end{subfigure}
    \caption{Evaluation of network latency impacts using the \texttt{train} scene and \texttt{NTHU/user3} trajectory.}
    \label{fig:rtt_evaluation}
\end{figure}

\subsubsection*{Impact of Round-Trip Time.}
To isolate the specific impact of core network latency, we executed targeted experiments varying the artificially injected Round-Trip Time (RTT). These evaluations were conducted rendering the \texttt{train} scene driven by the \texttt{NTHU/user3} movement trajectory of the EyeNavGS dataset. Figure~\ref{fig:rtt_evaluation}(a) illustrates the per-frame response time under a baseline \qty{20}{\milli\second} RTT scenario, while Figure~\ref{fig:rtt_evaluation}(b) summarizes the average response times across an incrementally scaling RTT environment. The findings indicate that the proposed Latency ABR registers a slightly higher overall response time compared to standard baseline ABRs. However, this marginal trade-off translates into definitive perceptual gains. 

\section{Conclusions}
In this work, we presented TIGAS, a novel and modular testbed for interactive 3DGS streaming over HTTP/3. Designed to serve as a versatile backbone for delivering spatial content when client devices lack the requisite power or bandwidth for autonomous rendering, TIGAS facilitates the rigorous evaluation of arbitrary 3DGS models, custom ABR logics, user movement traces, and neural enhancement modules. By offloading rasterization to the backend and employing frame-level ABRs, TIGAS decouples visual fidelity from client hardware, enabling real-time, 6DoF navigation on commodity devices, with an average quality of 0.88 SSIM. 

\balance
\bibliographystyle{plain}
\bibliography{ref}

\end{document}